\documentclass[11pt,letterpaper]{article}
\usepackage[letterpaper, margin=1in]{geometry}

\usepackage{graphicx,color}
\usepackage[cp850]{inputenc}
\usepackage[T1]{fontenc}
\usepackage{rotating}
\usepackage[f]{esvect}
\usepackage{amsmath, amsthm, amssymb}
\usepackage{rotating}
\usepackage{lscape}
\usepackage{mathrsfs}
\usepackage{subcaption}
\usepackage{mathtools}
\usepackage[toc,page]{appendix}
\usepackage{hyphenat}
\usepackage{mathptmx}
\usepackage[scaled=.65]{helvet}
\usepackage{courier}
\usepackage[toc,page]{appendix}
\usepackage{authblk}

\usepackage[normalem]{ulem}

\def\tablenotes{\bgroup\parfillskip=0pt plus 1fil
\leftskip=0pt\relax \rightskip=0pt
\vskip2pt\footnotesize}
\def\endtablenotes{\vskip1pt\egroup}

\newtheorem{theorem}{Theorem}[section]

\newtheorem{remark}[theorem]{Remark}

\renewcommand{\epsilon}{\varepsilon}
\renewcommand{\leq}{\leqslant}
\renewcommand{\geq}{\geqslant}
\renewcommand{\d}{\mathrm{d}}
\newcommand{\Diff}{\mathrm{D}}
\newcommand{\Abar}{\bar{A}}
\newcommand{\Bbar}{\bar{B}}

\renewcommand{\epsilon}{\varepsilon}

\newcommand{\TimeDeriv}{\frac{\textrm{d}}{\textrm{dt}}}

\allowdisplaybreaks

\usepackage[square,authoryear,sort]{natbib}

\usepackage{graphicx}
\usepackage{overpic}
\usepackage{tikz}
 \graphicspath{ {./Figures/} }
 
\begin{document}

\title{A continuation technique for maximum likelihood estimators in biological models}
 \author[1,*]{Tyler Cassidy}
 \affil[1]{School of Mathematics, University of Leeds, Leeds, LS2 9JT, UK}
 \affil[*]{t.cassidy1@leeds.ac.uk}

\maketitle
\section*{Abstract} 

Estimating model parameters is a crucial step in mathematical modelling and typically involves minimizing the disagreement between model predictions and experimental data.
This calibration data can change throughout a study, particularly if modelling is performed simultaneously with the calibration experiments, or during an on-going public health crisis as in the case of the COVID-19 pandemic. Consequently, the optimal parameter set, or maximal likelihood estimator (MLE), is a function of the experimental data set. Here, we develop a numerical technique to predict the evolution of the MLE as a function of the experimental data. We show that, when considering perturbations from an initial data set, our approach is significantly more computationally efficient that re-fitting model parameters while resulting in acceptable model fits to the updated data. We use the continuation technique to develop an explicit functional relationship between fit model parameters and experimental data that can be used to measure the sensitivity of the MLE to experimental data. We then leverage this inverse sensitivity analysis to select between model fits with similar information criteria, \textit{a priori} determine the experimental measurements to which the MLE is most sensitive, and suggest additional experiment measurements that can resolve parameter uncertainty. 

 \clearpage
\section{Introduction}

As quantitative modeling becomes more prevalent across biology and medicine \citep{Altrock2015,Perelson2002,Sanche2020}, mathematical models are increasingly being developed during the experimental data collection that will inform model parameters. This cooperation facilitates the use of mathematical modelling to inform experimental design and suggest potential intervention strategies \citep{Zhang2022,Sanche2020,Cardenas2022,Luo2022}. The COVID-19 pandemic is a striking example of the resulting feedback loop, where mathematical models suggest intervention strategies that influence the evolving public health crisis before being re-calibrated to new data. \citep{Holmdahl2020,Thompson2020,Davies2020}. 

Each updated data set requires re-calibration of the model typically through computationally expensive optimization techniques. To reduce this computational cost of the re-calibration step, it is common to use the existing parameters as a starting point when performing parameter fitting to incoming experimental data sets. This approach recycles optimization work but does not utilize leverage the relationship between the initial and updated experimental data set.  Here, we present a computational method to incorporate information about evolving data sets during the model validation and parameter estimation steps.

Specifically, for given model parameters and an initial experimental data set, we develop a method to predict the best-fit parameter set to an updated experimental data set. Our approach can be viewed as a numerical continuation technique \citep{Dhooge2008,DeSouza2019}. However, rather than studying the dynamical properties of the mathematical model as a function of model parameters, we consider the evolution of best-fit model parameters as a function of the experimental data.  We use the necessary condition for a local optima to write the best-fit parameters as an implicit function of the experimental data. Thus, we predict best-fit parameter sets for evolving experimental data without performing any optimization. Avoiding optimization leads to significant computational savings and we demonstrate these gains via two examples. In both these examples, our prediction method produces comparable model fits to randomly perturbed data sets to optimization techniques without the computational cost of solving the inverse optimization problem. 

While our approach does lead to increased computational efficiency, the more immediate application of our work may be in experimental design. Specifically, we identify an explicit relationship between individual best-fit parameter values and individual experimental data points through our continuation approach. We can therefore quantify which experimental measurements are the most informative for determining best-fit parameters and measure the sensitivity of parameter estimates to perturbations in data. The role of experimental design in model selection and parameterization has been extensively studied \citep{Silk2014,Cardenas2022,Li2015a,Li2013}. In particular, \citet{Li2015a} studied how correlations between best-fit model parameters can impact practical and structural identifiability of model parameters while \citet{Silk2014,Cardenas2022} explored how experimental design impacts model selection from a class of possible mathematical models. Conversely, our contribution explicitly relates individual experimental measurements with individual best-fit parameter estimates. We explicitly link our continuation technique to the Fisher information matrix commonly used in optimal experimental design \citep{Kreutz2009,Braniff2019}. Taken together, our approach allows the increased confidence in model parametrization from optimal experimental design to be mapped directly to individual model parameters. Accordingly, we can therefore design experiments to address specific uncertainties in parameter estimates. 

Furthermore, our work offers a distinct step towards understanding how robust parameter estimates are to evolving data. Many existing computational methods quantify confidence in parameterization; formal parameter sensitivity analyses \citep{Marino2008,Maiwald2016,Zi2011}, virtual population approaches \citep{Allen2016,Cassidy2019,Jenner2020}, or parameter identifiability analysis \citep{Castro2020}, often via profile likelihood computation \citep{Raue2009,Raue2014,Kreutz2012}, quantify how robust model predictions are to parameter variation. In particular, these techniques view the experimental data as fixed up to experimental noise and focus on the relationship between model parameters and model predictions. We offer a complementary approach to existing sensitivity analysis by explicitly studying how the best-fit parameters vary due to changes in calibration data. As we will see, our approach encodes information from local sensitivity analysis when calculating the functional relationship between the best-fit parameters and the calibration data. Consequently, while classical sensitivity analysis quantifies variability in model output due to change in model parameters, our approach considers changes in model parameters, and thus model predictions, as a function of the calibration data. We demonstrate this mapping of experimental data to best-fit parameter via an example drawn from mathematical oncology \citep{Cassidy2021}. These results, when combined with existing information criteria like the AIC or BIC \citep{Kass1995}, allow for modellers to quantify the robustness of best-fit parameter estimates when comparing different model fits to experimental data. 

The remainder of the article is structured as follows. We begin by defining the optimization problem in Section~\ref{Sec:Formulation}. We develop the continuation method in Section~\ref{Sec:ContinuationTechnique}, discuss our numerical implementation in \ref{Sec:NumericalImplementation}, and explore the connection between our continuation approach and classical profile likelihood in \ref{Sec:IdentifiabilityRelationship}. We then turn to two examples from mathematical biology to illustrate the utility of our technique in Section~\ref{Sec:Examples} before finishing with a brief discussion.

\section{Methods}

\subsection{Formulation of the optimization problem}\label{Sec:Formulation}
Here, we introduce the framework of the underlying optimization problem. We focus on ordinary differential equation (ODE) models representing biological processes, as these models are common throughout mathematical biology. However, our approach extends to partial differential equation or delay differential equation models directly. We consider a generic ODE based model throughout the remainder of this work.

Let the model states be given by $x(t) \in \mathbb{R}^n$ with model parameters denoted by $\theta \in \Omega \subset \mathbb{R}^p$ where $\Omega$ is a subset of biologically plausible parameter values. We explicitly allow the initial condition $x(0)$ to depend explicitly on the model parameters $\theta$. Taken together, we consider the differential equation model
\begin{align}
\TimeDeriv x(t) = f(x,\theta); \quad x(0) = x_0(\theta)
\label{Eq:GenericModel}
\end{align}
where $f$ is continuously differentiable in $x$ and $\theta$. 

We consider calibration data $\{ \phi_i \}_{i=1}^{d \times m}$ representing $m$ measurements each taken at $d$ time points $\{ t_i \}_{i=1}^d$. It is possible that model species are not directly comparable against the calibration data so we define the $m$ model observables by 
\begin{align*}
y_i(\theta) = h(x(t_i,\theta),\theta) \in \mathbb{R}^{d \times m}.
\end{align*}
In what follows, we consider $m=1$ for notational simplicity although the analysis extends for $m \geq 2$.   

\subsubsection*{Likelihood function and objective function}

\begin{remark}
The methods that follow do not assume a specific objective function. However, we do assume that the objective function is twice continuously differentiable as is commonly the case. For simplicity, we present the remainder of our results using the common log-likelihood formulation \citep{Stapor2018,Maiwald2016}.
\end{remark}

The likelihood describes the probability of observing experimental data $\phi$ as a function of $\theta$ and is given by
\begin{align}
\mathcal{L}(y(x(t,\theta)),\phi) =  \prod_{i=1}^d \frac{1}{\sqrt{ 2\pi \sigma^2_{i} }} \exp\left[ -\frac{(y_i(\theta)- \phi^*_i)^2}{\sigma_i^2} \right]
\label{Eq:Likelihood}
\end{align}
The experimental error at each measurement point, $\sigma_i$, can be estimated as an additional model parameter or fixed to a known value. Here, we follow \citet{Sharp2022} and take $\sigma_i$ fixed at a known constant value, although it is possible to include $\sigma_i$ in the vector of unknown parameters $\theta$. The maximum likelihood estimator (MLE) $\theta^*$, and thus best-fit model parameters for the given experimental data $\phi$, is defined by the solution of the inverse problem
\begin{align*}
\theta^* = \textrm{argmax}_{\theta \in \Omega} \mathcal{L}(\theta,\phi^*).
\end{align*}
As the differential equations defining $y(x(t,\theta))$ rarely have explicit solutions, the likelihood \eqref{Eq:Likelihood} is difficult to evaluate analytically. It is therefore standard to minimize the negative log-likelihood $G(\theta,\phi) = - \log \left( \mathcal{L}(y(x(t,\theta)),\phi^*) \right)  $ given by  
\begin{align}
G(\theta,\phi) = \displaystyle \sum_{i=1}^d \log\left( \sqrt{2\pi \sigma_i^2 } \right) + \frac{(y_i(\theta)- \phi_i^*)^2}{\sigma_i^2}.  
\label{Eq:LogLikelihood}
\end{align}
Under the assumption that $\sigma_i = \sigma$ is fixed, the error term $\log\left( \sqrt{2\pi \sigma^2 } \right)$ and denominator of $G(\theta,\phi)$ are constant and do not influence the solution of the optimization problem. The maximum likelihood estimator $\theta^*$ is the parameter set that minimizes $G(\theta,\phi^*)$.  A number of computational techniques exist to minimize $G(\theta,\phi)$ and thus calculate $\theta^*$. These optimization techniques typically require simulating the mathematical model \eqref{Eq:GenericModel} at each optimization step. Further complicating the optimization, $G(\theta,\phi^*)$ is often non-convex with multiple local minima.

\subsection{Continuation of maximal likelihood estimator} \label{Sec:ContinuationTechnique}

In \eqref{Eq:LogLikelihood}, we explicitly write the objective function $G$ as a function of the model parameters $\theta$ and the experimental data $\phi$. Accordingly, the MLE $\theta^*$ is an implicit function of the experimental data $\phi$ defined as the solution of the optimization problem
\begin{align}
\theta^*(\phi) = \textrm{argmax}_{\theta \in \Omega} \mathcal{L}(\theta,\phi).
\label{Eq:MLEDefn}
\end{align}
Model fitting is increasingly performed concurrently with experiments \citep{Luo2022} or obtained from an evolving real-world scenario, as in epidemic modelling \citep{Sanche2020}. In both of these cases, the experimental data is evolving and should not be considered as known and constant. Accordingly, we are interested in the MLE as a function of the experimental data $\phi$. Most existing optimization techniques consider the experimental data fixed and omit this dependence. Here, we develop a continuation type technique to compute the evolution of $\theta^*$ numerically as a function of $\phi$ from an initial solution of the optimization problem. Ultimately, we calculate the evolution of $\theta^*(\phi)$ as the calibration data varies to generate a curve of potential MLEs in $(\phi,\theta^*)$ space using a numerical continuation technique.

Numerical continuation methods compute branches of implicitly defined curves. A standard application of these continuation type techniques in mathematical biology is numerical bifurcation analysis \citep{Dhooge2008,Sanche2021}. In their most common form, numerical bifurcation techniques compute equilibrium systems of a non-linear dynamical system as a function of model parameters but can be used to detect much richer dynamical behaviour \citep{DeSouza2019}. Often, these continuation techniques leverage ``predictor-corrector'' algorithms. Predictor-corrector approaches use the implicit function theorem to predict the solution to the corresponding non-linear system of equations. Then, the predicted solution is used as a starting value to explicitly calculate the solution of the system of equations during the corrector step. Here, we develop a similar ``prediction-correction'' strategy to predict the behaviour of the solution $\theta^*(\phi)$ of the inverse problem~\eqref{Eq:MLEDefn} as a function of the data $\phi$. We focus on the ``predictor'' step, as the corrector step, if necessary, can utilize existing numerical optimization techniques to calculate the MLE. 

As the log-likelihood \eqref{Eq:LogLikelihood} is continuously differentiable, local optimal must satisfy 
\begin{align}
\Diff_\theta G(\theta^*,\phi) = 0,
\label{Eq:ContinuationEqn}
\end{align}
so we necessarily have 
\begin{align*}
\theta^*(\phi) \in \{ \theta \in \Omega | \Diff_{\theta} G(\theta^*,\phi) = 0 \}. 
\end{align*}
However, unlike the implicit equation used to determine equilibria of a dynamical system and used in continuation techniques for numerical bifurcation analysis, the optimality condition \eqref{Eq:ContinuationEqn} is a necessary, but not sufficient, condition for $\theta^*$ to be a MLE. Models that are not structurally identifiable \citep{Raue2014} have manifolds in parameter space on which this optimality constraint holds but are not necessarily MLEs.  We discuss the relationship between our approach and profile likelihood classifications of structural identifiability in Section~\ref{Sec:IdentifiabilityRelationship}. 

Now, let $\theta^*_0$ be the MLE for calibration data $\phi_0$. Further, let the Hessian $\Diff^2_\theta G(\theta,\phi)$  be invertible at $ (\theta_0^*,\phi_0) \in \mathbb{R}^p \times \mathbb{R}^d$ and consider the function
\begin{align*}
\Diff_{\theta} G(\theta^*,\phi): \mathbb{R}^p \times \mathbb{R}^d \to \mathbb{R}^p.
\end{align*}
Then, the implicit function theorem ensures the existence of a function $\Psi(\phi)$ such that 
\begin{align*}
\Diff_{\theta} G(\Psi(\phi) ,\phi) = 0
\end{align*} 
in a neighbourhood of $\phi_0$ with $\Psi(\phi_0) = \theta^*(\phi_0)$. It is natural to consider $\Psi(\phi)$ as the predicted MLE $\theta^*(\phi)$ for $\phi$ in a neighbourhood of $ \phi_0$.

The implicit function theorem ensures that $\Psi$ exists but computing $\Psi(\phi)$ analytically is functionally impossible. However, the implicit function $\Psi(\phi)$ is continuously differentiable and we expand $\Psi$ as a function of the calibration data $\phi$ using Taylor series 
\begin{align}
\Psi(\phi+\Delta \phi) = \Psi(\phi) + \Diff \Psi(\phi)\Delta \phi + \mathcal{O}(\Delta \phi ^2).
\label{Eq:PredictorEquation}
\end{align}
where $\phi + \Delta \phi$ is the updated calibration data. Then, to predict $\Psi$ starting from a known solution $\Psi(\phi) = \theta^*$ we calculate $\Diff \Psi(\phi)$. The implicit function theorem implies that
\begin{align*}
\Diff \Psi = - \left[\Diff_{\theta}^2 G(\Psi(\phi),\phi)\right]^{-1} \Diff_{\theta,\phi}^2 G(\Psi(\phi),\phi).
\end{align*}
We thus use $\Diff \Psi$ to evaluate \eqref{Eq:PredictorEquation} and thus perform the continuation step. 


\subsection{Numerical Implementation} \label{Sec:NumericalImplementation}

We now show how to use the objective function \eqref{Eq:LogLikelihood} to calculate finite difference approximations to the derivatives included in \eqref{Eq:PredictorEquation}. As before, we assume that we are given a point $(\theta_0^*,\phi_0) \in \mathbb{R}^p \times \mathbb{R}^d$ such that 
\begin{align*}
\theta_0^* = \textrm{argmin}_{\theta \in \Omega} G(\theta,\phi_0).
\end{align*}

For $\theta_n$ denoting the $n$-th parameter, we calculate 
\begin{align*}
 \frac{\partial G(\theta,\phi) }{\partial \theta_n} = \displaystyle \sum_{i=1}^d 2\left( y_i(\theta) - \phi_i\right) \frac{ \partial y_i(\theta)}{\partial \theta_n}
\end{align*}
and so
\begin{align}
\left[ \Diff_{\theta,\phi}^2 G(\Psi(\phi),\phi)\right]_{(n,i)} =  -2 \frac{ \partial y_i(\theta)}{\partial \theta_n}. 
\label{Eq:MixedPartialExpression}
\end{align}
The derivatives $\partial_{\theta_n} y_i(\theta)$ can be calculated through finite difference schemes \citep{Zi2011}  
\begin{align*}
\frac{ \partial y_i(\theta)}{\partial \theta_n} = \frac{y_i(\theta+\Delta \theta_n) - y_i(\theta-\Delta \theta_n)  }{2\Delta \theta_n} + \mathcal{O}\left( (\Delta \theta_n)^2 \right) ,
\end{align*} 
where $\Delta \theta_n$ is a small perturbation in only the $n$-th parameter. In practice, it is standard to take $\Delta \theta_n$ to be some small percentage of the initial parameter $\theta_n$ \citep{Li2011}. In this case, computing $\Diff_{\theta,\phi}^2 G(\Psi(\phi),\phi)$ requires $2p$ model simulations where $p$ is the number of model parameters. We note that $\partial_{\theta_n} y_i(\theta)$ is commonly used to perform local sensitivity analysis and that more accurate finite difference approximations, such as centered differences, can be used to calculate $\Diff_{\theta,\phi}^2 G(\Psi(\phi),\phi)$. 

Calculating the Hessian $\Diff_{\theta}^2 G(\theta,\phi)$ via finite differences is simple to implement but computationally expensive due to the number of objective function evaluations. However, the Hessian, or the observed Fisher Information, is commonly used throughout parameter optimization algorithms and other techniques such as profile likelihood calculations, estimates of the likelihood function, and classical sensitivity anaylsis, which has led to recent advances in the development of computationally efficient techniques to calculate $\Diff_{\theta}^2 G(\theta,\phi)$ \citep{Stapor2018} and the ability to recycle these calculations to avoid computational cost.

In the following examples, we use a  finite difference scheme to calculate $\Diff_{\theta}^2 G(\theta,\phi)$. We calculate the diagonal elements of  $\Diff_{\theta}^2 G(\theta,\phi)$ using forward second order differences and the off-diagonal terms by 
\begin{align*}
\frac{\partial G(\theta,\phi)}{\partial \theta_i \partial \theta_j} & = \left( \frac{1}{4(\Delta \theta_i) (\Delta \theta_j)} \right) \left[  G(\theta+\Delta \theta_i+\Delta \theta_j,\phi) - G(\theta+\Delta \theta_i - \Delta \theta_j,\phi) \right. \\
& {} \quad \left.  + G(\theta - \Delta \theta_i+\Delta \theta_j,\phi)+ G(\theta-\Delta \theta_i-\Delta \theta_j,\phi) \right] + \mathcal{O}\left( (\Delta \theta_i)^2, (\Delta \theta_j)^2\right). 
\end{align*}
Thus, our computation of the Hessian requires $2p(p+1)$ objective function evaluations, although, as mentioned, more efficient implementations are available. In fact, many gradient-based optimization techniques approximate the Hessian $D^2_{\theta,\theta} G(\theta,\phi)$ at each iteration \citep{Matlab2017}. For example, both \texttt{fmincon} and \texttt{fminunc} in \citep{Matlab2017} calculate $D^2_{\theta,\theta} G(\theta,\phi)$ at each step and print the pre-computed Hessian as an output of the optimizer. It is therefore possible, and efficient, to recycle this calculation when calculating an update to $ \theta_0^*$ using \eqref{Eq:ContinuationEqn}. 

All told, this numerical implementation requires $2p(p+2)$ model simulations to evaluate \eqref{Eq:ContinuationEqn}. This computational cost is certainly not optimal but does benefit from re-using calculations performed in local sensitivity analysis and the optimization step. Finally, while we have written \eqref{Eq:ContinuationEqn} with the inverse of $\Diff_{\theta}^2 G(\theta,\phi)$, it is computationally more appropriate to solve the linear system of equations
\begin{align*}
\Diff_{\theta}^2 G(\theta,\phi) \Diff \Psi = -   \Diff_{\theta,\phi}^2 G(\Psi(\phi),\phi)
\end{align*}
for the unknown $\Diff \Psi$. 

Code to implement this continuation technique is available at \texttt{https://github.com/ttcassid/\allowbreak MLE$\_$Continuation}.

\section{Results}
 
\subsection{Relationship with existing techniques}\label{Sec:IdentifiabilityRelationship}

There are a number of existing techniques to study the relationship between model parameters and data. While our continuation technique focuses on the relationship between the MLE and the calibration data, it has many ties to these existing techniques. We therefore discuss how this continuation method relates to parameter identifiability as assessed by the profile likelihood; local sensitivity analysis; and experimental design, with a focus on using the explicit relationship between data and the MLE to suggest additional experimental measurements. 

\subsubsection*{Parameter identifiability}
Thus far, we have explicitly written the MLE estimator as a function of the experimental data used  to fit a model. Our approach is intrinsically related to parameter identifability analysis. Identifiability analysis attempts to determine if available experimental observations are capable to uniquely determine  model parameters. Accordingly, the practical identifiability of a mathematical model depends on available experimental data. The \textit{profile likelihood}, given by  
\begin{align*}
PLE_{\theta_i}(c) = \min_{\theta_i = c, \theta \in \mathbb{R}^p } G(\theta,\phi),
\end{align*}
and introduced by \citet{Raue2009}, is a projection of the likelihood function onto the model parameter $\theta_i = c$. The profile likelihood illustrates the behaviour of the likelihood function as the parameter $\theta_i$ is fixed away from the optimal value $\theta^*_i$. The shape of $PLE_{\theta_i}(c)$ illustrates the confidence interval of the parameter estimate $\theta_i^*$ for given experimental data. Formally, \citet{Raue2009} define these confidence intervals by
\begin{align*}
\textrm{C.I.}(\theta_i,\alpha) = \{ c | PLE_{\theta_i}(c) - PLE_{\theta_i}(\theta_i^*) < \Delta_{\alpha} \}
\end{align*}
where $\Delta_{\alpha} = \chi^2(\alpha,df)$ is the $\chi^2$ distribution at significance level $\alpha$ and $df$ degrees of freedom \citep{Raue2009}. A parameter is practically identifiable in the sense of \citet{Raue2009} with confidence level $\alpha$ if $\textrm{C.I.}(\theta_i,\alpha)$ is bounded in parameter space for given experimental data. Conversely, a non-identifiable parameter has a profile likelihood that does not increase past the threshold $\Delta_{\alpha}$.

The profile likelihood is intrinsically linked to the available experimental data $\phi_i$. We view the PLE as a function of both the parameter $\theta_i$ and the experimental data $\phi$
\begin{align*}
PLE_{\theta_i}(c,\phi) = \min_{\theta_i = c, \theta \in \mathbb{R}^p } G(\theta,\phi).
\end{align*}
For practically unidentifiable models, it is natural to ask what perturbations to the experimental data could render the model practically identifiable. \citet{Raue2009} use the profile likelihood of a model parameter to suggest additional experiments to resolve practical non-identifiability. They simulate the model for parameter values along $PLE_{\theta_i}$ to suggest additional experimental measurements at times $t_{s,i}$, where $t_{s,i}$ represents the $i-$th \textit{simulated} measurement time. In our framework, we define 
\begin{align*}
\theta^*|_{\theta_i = c }(\phi) = \textrm{argmin}_{\theta_i = c, \theta \in \mathbb{R}^p}  G(\theta,\phi),
\end{align*}
so that 
\begin{align*}
PLE_{\theta_i}(c,\phi) =  G(\theta^*|_{\theta_i = c }(\phi) ,\phi).
\end{align*}
We note that the definition of $\theta^*|_{\theta_i = c }(\phi)$ is precisely that of $\theta^*(\phi)$ with the added constraint that $\theta_i = c$. We can calculate $\Diff_{\phi} \theta^*|_{\theta_i = c }$ as a function of the experimental data $\phi$ in precisely the same manner as described previously. Consequently, our continuation approach can complement the experimental design approach suggested by \citet{Raue2009} by incorporating the sensitivity of the MLE to perturbations in the (simulated or experimental) calibration data.


\subsubsection*{Sensitivity analysis} \label{Sec:LocalSensitivity}

Local sensitivity analysis quantifies how small perturbations of the best-fit parameters impact model output \citep{Zi2011}. A standard approach to local sensitivity analysis is using the finite difference approximation of
\begin{align*}
S_n(t) = \frac{ \partial y (\theta)}{\partial \theta_n} =  \frac{h(t_i,\theta+\Delta \theta_n) - h(t_i,\theta-\Delta \theta_n)  }{\Delta \theta_n} + \mathcal{O}\left( \Delta \theta_n \right)  
\end{align*} 
to identify which parameter values strongly impact model projections. When $|S_n|$ is small, the model output is considered to be insensitive to $\theta_n$. The $n$-th row of $\Diff_{\theta,\phi}^2 G(\Psi(\phi),\phi)$ is precisely $S_n(t_i)$ for $t_i$ corresponding to calibration data measurements.   
When implementing \eqref{Eq:ContinuationEqn}, the magnitude of the continuation step $\Diff \Psi(\phi)\Delta \phi$ in the direction of $\theta_n$ is scaled by $S_n$. This scaling encodes the local sensitivity of model predictions to variations in parameters in the prediction of $\Psi(\phi)$. Consequently, our continuation method naturally includes the information gained from local sensitivity analysis.

\subsubsection*{Experimental design}

In our derivation of $\Diff \Psi$, we assumed that the Hessian matrix $\Diff_{\theta}^2 G(\theta,\phi)$ was invertible. The Hessian gives the curvature of the loglikelihood and is known as the observed Fisher information matrix $\mathcal{I}_{obs}$. The observed Fisher information is a local measurement in data space. Conversely, the expected Fisher information considers the entirety of data space for fixed model parameters $\theta$.  The expected Fisher information is obtained by taking the expectation of $\Diff_{\theta}^2 G(\theta,\phi)$ over all possible experimental measurements $\phi$ and is defined via
\begin{align*}
\mathcal{I} = \mathbb{E}\left[\Diff_{\theta}^2 G(\theta,\phi)\right].
\end{align*} 
Many existing experimental design methods leverage the expected Fisher information matrix to minimize the covariance in model parameter estimates via the Cram\'{e}r-Rao inequality. These experimental design techniques typically maximize some aspect, often the determinant, of the Fisher information matrix as a function of possible data to select the most informative calibration data set \citep{Kreutz2009}. From a geometric perspective, maximizing the determinant of the Fisher information matrix corresponds to minimizing the volume of the confidence ellipsoid engendered from the covariance matrix \citep{Braniff2019}.

In particular, \citet{Braniff2019a} considered the case of bistable gene regulatory networks where the fold bifurcation and unstable manifold between stable equilibria complicates experimental design and parameter estimation.  \citet{Sharp2022} considered an information-geometry perspective to propose the expected Fisher information matrix and resulting Riemannian manifold as a guide for data collection. As is often the case, both \citet{Sharp2022} and \citet{Braniff2019a} used the expected Fisher information, which considers all possible calibration data via the expectation over $\phi$. Here, we show how our approach complements the classical Fisher information approach to experimental design, albeit through a local measurement, in $(\theta,\phi)$ space. We recall that
\begin{align*}
\Diff \Psi\Delta \phi  = -\left[\mathcal{I}_{obs}\right]^{-1} \Diff_{\theta,\phi}^2 G(\Psi(\phi),\phi) \Delta \phi, 
\end{align*}
so if $\Diff_{\theta,\phi}^2 G(\Psi(\phi),\phi)$ were the identity, then $\Diff \Psi$ would correspond to the Fisher information approach to measuring uncertainty in MLE. 

In the calculation of $\Diff \Psi \Delta \phi$, the matrix $\Diff_{\theta,\phi}^2 G(\Psi(\phi),\phi)$ maps perturbations in the calibration data $\Delta \phi$ through the curvature of the loglikelihood  to changes in the MLE. Consequently, $\Diff_{\theta,\phi}^2 G(\Psi(\phi),\phi)$ acts as a change of basis matrix from the space of calibration data to parameter space. Simply, $\Diff_{\theta,\phi}^2 G(\Psi(\phi),\phi)\Delta \phi$ scales changes in the calibration data to the confidence ellipsoid in parameter space obtained from $\left[\mathcal{I}_{obs}\right]^{-1}$. Geometrically, if $\Diff^2_{\theta} G$ has eigenvalues $\lambda_i$ with corresponding eigenvectors $\nu_i$, then choosing $\Delta \phi$ such that $\nu_i = \Diff^2_{\theta} G\Delta \phi$ translates perturbations in calibration data to the corresponding eigenspace of the covariance matrix.

For example, the $i-$th column of $ \Diff \Psi $ maps perturbations of the $i-$th data point to changes in the MLE. Specifically, the sum
\begin{align*}
\frac{\Delta \theta^*}{\Delta \phi_k} = \displaystyle \sum_{k=1}^p | \Diff \Psi_{k,j} |
\end{align*}
measures the sensitivity of the MLE $\theta^*$ to perturbations in the $k-$th data point. Thus, 
\begin{align*}
\| \Diff \Psi \|_{1} = \displaystyle \max_{k= 1,2,...,p} \frac{\Delta \theta^*}{\Delta \phi_k} 
\end{align*}
and the most informative data point satisfies 
\begin{align*}
l = \displaystyle \textrm{argmax}_{k= 1,2,...,p} \displaystyle \frac{\Delta \theta^*}{\Delta \phi_k} , 
\end{align*}
where informative is understood as the data point inducing the largest sensitivity in the MLE. As an extreme example, if 
\begin{align*}
\frac{\Delta \theta^*}{\Delta \phi_n} = 0,
\end{align*}
then perturbations in $\phi_n$ do not impact the MLE estimate, which implies complete insensitivity of the model fit to $\phi_k$. This example corresponds to $\Delta \phi$ belonging to the kernel of the matrix $\Diff^2_{\theta,\phi}G$ since we have assumed that $\Diff^2_{\theta}G$ is invertible. 

We can therefore utilize our analysis to identify which additional experimental measurements could increase confidence in model parameterization. Consider $k$ additional measurements $\{\phi_{s,i}  = y_{s,i}(\theta^*)\}_{i=1}^k$ taken directly from the model simulation at times $\{t_{s,i}\}_{i=1}^k$ where the subscript $s$ indicates simulated data. Including $\{\phi_{s,i}\}$ in the objective function~\eqref{Eq:LogLikelihood} does not change the MLE or objective value function as these simulated data exactly match the model values. However, $\|\Diff \Psi (\phi+ \Delta \phi_{s,i})\|$ quantifies the sensitivity of the MLE to variability in the $k$ simulated measurements.  Accordingly, the measurement that maximizes $\|\Diff\Psi (\phi+ \Delta \phi_{s,i})\|$ for a fixed perturbation size $\Delta$ is a good candidate for an additional experimental measurement to decrease parameter uncertainty. 

\subsection{Examples}\label{Sec:Examples}
 
The continuation framework derived earlier is applicable to a large variety of models throughout in the mathematical biology literature. To demonstrate the utility of the continuation method, we consider two examples from distinct fields and model formulations. First, we consider a mathematical model of phenotypic heterogeneity in non-small cell lung cancer (NSCLC) \citep{Cassidy2021}. This model is given by a system of two non-local, structured PDEs representing the density of drug-sensitive and drug-tolerant NSCLC cells. The PDE model is equivalent to a system of integral equations following the introduction of two auxiliary variables which can be further reduced to a system of ODEs (see \citep{Cassidy2021} for details). The parameters of the ODE model were fit to \textit{in vitro} NSCLC data taken from growth experiments in treated and untreated media \citep{Cassidy2021}. 

We also consider a classical model of HIV-1 viral dynamics. This model has been used extensively to understand viral dynamics data \citep{Perelson2002} and the identifiability of model parameters was considered by \citet{Wu2008}. In that work, \citet{Wu2008} used simulated data to validate their identifiability results; we follow \citet{Wu2008} and use simulated data to illustrate our approach.
 
\subsubsection*{A PDE model of phenotypic switching in mathematical oncology} \label{Sec:PhenotypeExample}

Non-genetic phenotypic heterogeneity has been increasingly studied as a driver of treatment resistance in solid cancers \citep{Goldman2015}. A number of mathematical models have been derived to study the emergence of phenotypic plasticity in cancer cell lines \citep{Gunnarsson2020,Jolly2018,Sahoo2021,Craig2019}. We consider the \citet{Cassidy2021} model that tracks the density of NSCLC cells with a drug-sensitive ($A(t,a)$) or drug-tolerant ($B(t,a)$) phenotype at time $t$ and age $a$. The total number of cells of each phenotype is given by 
\begin{equation}
\Abar(t) = \int_0^{\infty} A(t,a)\d a \quad \textrm{and} \quad \Bbar(t) = \int_0^{\infty} B(t,a) \d a.
\label{Eq:AbarDefn}
\end{equation} 
The total number of NSCLC cells is given by $N(t) = \Abar(t) + \Bbar(t)$. \citet{Cassidy2021} considered logistic growth with an Allee effect, wherein cooperation between cells of the same phenotype can lead to increased growth rates, given by 
\begin{align} \notag
R_A(\Abar(t),\Bbar(t)) & = r_A\left(1-\frac{\Abar(t)+\Bbar(t)}{K} \right) \quad \textrm{and} \\
R_B(\Abar(t),\Bbar(t)) & = r_B \left(1-\frac{\Abar(t)+\Bbar(t)}{K} \right)f_n(\Abar(t),\Bbar(t)).
\label{Eq:LogisticGrowthRatesSI}
\end{align}
where $r_A$ and $r_B$ are phenotype specific growth rates, the carrying capacity is $K$, and the strength of the Allee effect is
\begin{align*}
f_n(\Abar(t),\Bbar(t)) = 1 +\left( \frac{r_A-r_B}{r_B}\right) \left(\frac{\Bbar(t)^n}{\Abar(t)^n+\Bbar(t)^n } \right).
\end{align*}
Finally, drug-tolerant and drug-sensitive cells have phenotype-specific death rates $d_B$ and 
\begin{equation*}
d_A = \left \{ 
\begin{array}{cc}
d_A & \textrm{If untreated} \\
d_A^{max} & \textrm{During treatment.}
\end{array}
\right.
\end{equation*}
$A(t,a)$ and $B(t,a)$ satisfy the age structured PDEs
\begin{equation}
\left.
\begin{aligned}
\partial_t A(t,a) + \partial_a A(t,a) & = -[d_A+R_A(\Abar(t),\Bbar(t))]A(t,a) \\
\partial_t B(t,a) + \partial_a B(t,a) & = -[d_B+R_B(\Abar(t),\Bbar(t))]B(t,a)
\end{aligned}
\right \}
\label{Eq:PhenotypePDE}
\end{equation}
with boundary conditions corresponding to cellular reproduction given by
\begin{equation} \label{Eq:BoundaryReproductionCondition}
\left. 
\begin{aligned}
A(t,0) & = 2  \hspace{-2pt}\int_0^{\infty}  \hspace{-10pt} \left[  R_A(\Abar(t),\Bbar(t)) \beta_{AA}(a)A(t,a) + f_n(\Abar(t),\Bbar(t))R_B(\Abar(t),\Bbar(t))\beta_{BA}(a)B(t,a) \right]  \d a\\ 
B(t,0) & = 2  \hspace{-2pt}\int_0^{\infty}  \hspace{-10pt} \left[ R_A(\Abar(t),\Bbar(t)) \beta_{AB}(a)A(t,a) + f_n(\Abar(t),\Bbar(t))R_B(\Abar(t),\Bbar(t))\beta_{BB}(a)B(t,a) \right] \d a .
\end{aligned}
\right \}
\end{equation} 
The functions $\beta_{ij}$ represent the probability of a reproducing mother cell with age $a$ and phenotype $i$ giving birth to a daughter cell with phenotype $j$. The probability of phenotypic inheritance is given by  
\begin{equation*}
\beta_{ii}(a) = P_{ii}^* + (P_{ii}^{max}-P_{ii}^*)\exp\left(-\sigma_{i} a\right),
\end{equation*}
where $\sigma_i$ represents the decay rate of intracellular signalling factors that modulate how ageing impacts the probability of daughter cells retaining the mother cells phenotype, and 
\begin{equation*}
\beta_{AB}(a) = 1-\beta_{AA}(a) \quad \textrm{and} \quad \beta_{BA}(a) = 1-\beta_{BB}(a). 
\end{equation*}
Further details, including a derivation of the initial conditions of \eqref{Eq:PhenotypePDE}, model analysis, and reduction of the phenotype switching mode~\eqref{Eq:PhenotypePDE} to a system of ODEs can be found in \citet{Cassidy2021}. 

The model~\eqref{Eq:PhenotypePDE} was fit to \textit{in vitro} experimental data corresponding to NSCLC cell population growth in untreated and treated environments where treatment is applied from day $3$ onwards. The calibration data is 4 data points $\{ \phi_i \}_{i=1}^4$ collected at time $t_i = 0,2,4,6$ days in the control experiment, and two additional data points $\{ \phi_i \}_{i=5}^6$ collected on days $t_i = 4,6$ days during the treated experiment. As anti-cancer treatment is applied from day 3 on-wards and decreases the cancer cell population, we necessarily have $\phi_5 \leq \phi_3$ and $ \phi_6 \leq \phi_4$. We denote the experimental data used to parametrize the model by $\{ \phi_i^0\}_{i=1}^6$. The model output corresponding to the experimental measurements is thus 
\begin{align*}
y_i (\theta) = N(t_i,\theta),
\end{align*}
and the objective function is the standard sum of squares error given by 
\begin{align*}
G_{pheno}(\theta,\phi) = \sqrt{ \displaystyle \sum_{i=1}^6 \left( \log_{10}(N(t_i,\theta) - \log_{10}(\phi_i) \right)^2 }. 
\end{align*}
\citet{Cassidy2021} fit model parameters $ [r_A, r_B, d_A=d_B, d_A^{max}] $ to treated and untreated experimental data simultaneously for a number of cell lines. The MLE found by \citet{Cassidy2021} corresponds to $\theta^*(\phi^0) =  [0.4827, 0.3498, 0.7025, 0.4198]$. 

We perturbed the experimental data collected by \citet{Craig2019} with increasing amounts of Gaussian noise. We created 10 perturbed data sets $\{ \phi_i^j \}_{i=1}^6$ where the index $j = 1,2,...,10,$ denotes the $j$-th perturbed data set and the normally distributed noise with $\mu = 0$,  $\sigma^2 = 1$, and scaled such that
\begin{align*}
\| \log_{10}(\phi_i^j) - \log_{10}(\phi_i^*) \| = (0.05 + jh_{step} \times \left( 0.75-2 \times 0.05 \right) \left(\frac{2}{10(11)} \right) \|  \log_{10}( \phi_i^0 ) \|
\end{align*}
where $ h_{step} = 0.65/55$ was chosen such that $\| \log_{10}(\phi_i^{10}) - \log_{10}(\phi_i^0) \| = 0.75 \|  \log_{10}( \phi_i^0) \|.$ 

We enforce that this randomly perturbed data satisfies $\phi_5 \leq \phi_3$ and $ \phi_6 \leq \phi_4$. For each perturbed data set $\{ \phi_i^j \}$, we used the continuation method described in Section~\ref{Sec:ContinuationTechnique} to calculate
\begin{align}\label{Eq:PhenotypeContinuation}
\Psi(\phi^j) = \theta^*(\phi^{j-1} ) + \Diff \Psi(\phi^{j-1}) \Delta \phi + \mathcal{O}(\Delta \phi ^2).
\end{align}
 
\begin{figure} [h!]
\includegraphics[scale=0.9, trim= 50 500 0 0,clip]{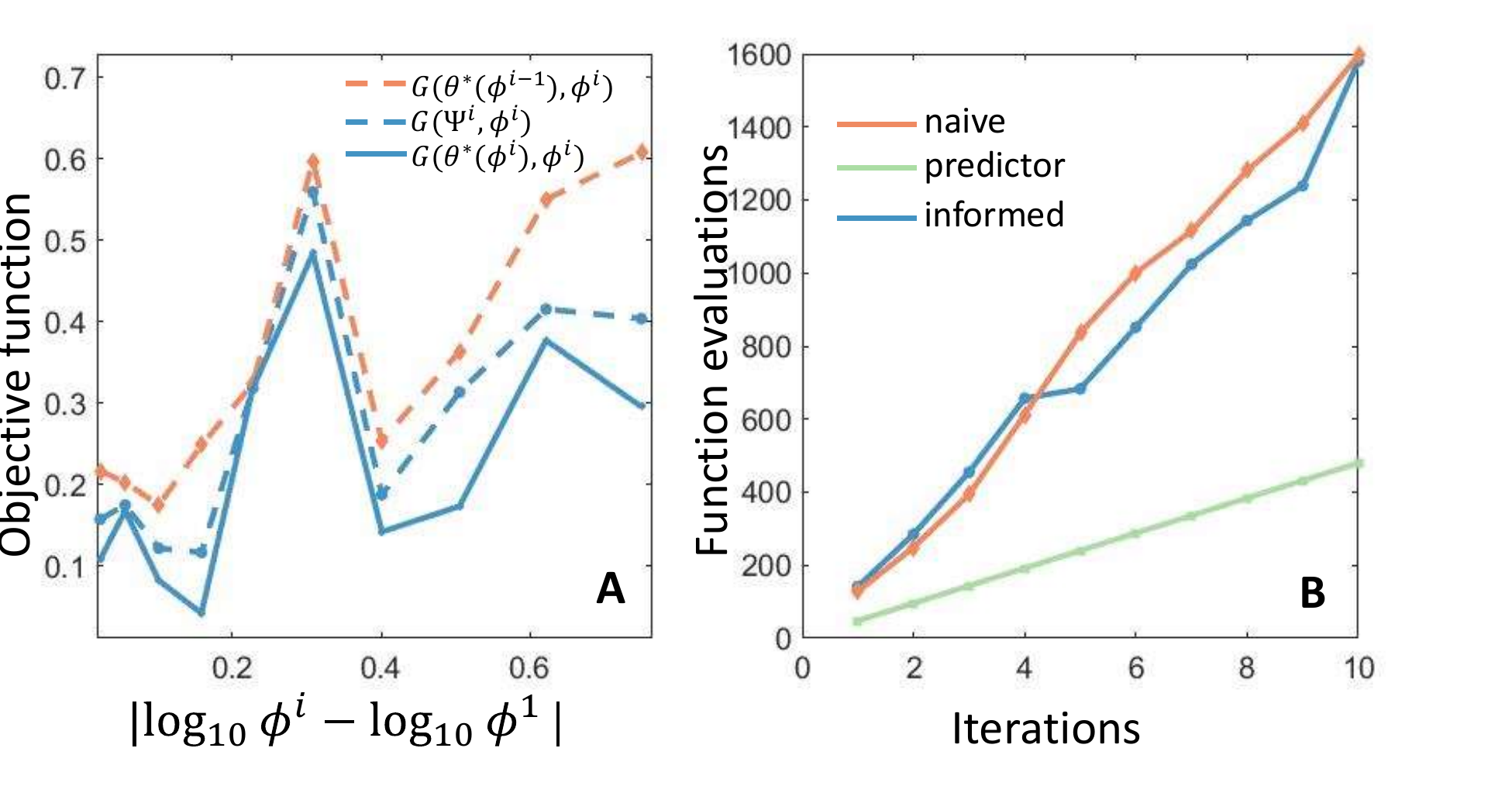} 
\caption{ Comparison between MLE estimates obtained using the naive and continuation approaches. Panel \textbf{A} shows a comparison of the objective function value for the naive and continuation guesses as well as the true minimal objective function value as a function of the perturbation of the experimental data from the initial data. Panel \textbf{B} shows a comparison of the number of objective value evaluations required to obtain the minimal value when starting from the naive or predicted MLE with the number of function evaluations required to calculate $\Psi(\theta^i)$. }
\label{Fig:PhenotypeExample}
\end{figure}

The naive approach to calculate the MLE $\theta^*(\phi^j)$ for updated data $\phi^j$ would be to use the MLE from the previous data, $\theta^*(\phi^{j-1})$, as an initial starting guess for the parameter fitting step.  Hence, to illustrate the utility of our continuation technique, we calculated $\Psi(\phi^j)$ using \eqref{Eq:PhenotypeContinuation} and then calculated $G_{pheno}(\Psi( \phi^j ),\phi^j)$. We also calculated the true MLE $\theta^*(\phi^j)$ using the Matlab algorithm \texttt{fmincon} from the starting guesses  $\Psi(\phi^j)$ and $\theta^*(\phi^{j-1} )$. In Figure~\ref{Fig:PhenotypeExample} A), we show the objective function value evaluated at the updated data $\phi^j$ and three parameter sets : the naive starting point, $\theta^*(\phi^{j-1})$; the predicted MLE, $\Psi(\phi^j)$; and the true MLE, $\theta^*(\phi^j)$.  We note that the non-monotonic profile of the objective function $G_{pheno}$ in Figure~\ref{Fig:PhenotypeExample} A) is to be expected as we are adding noise to experimental data. This noise may perturb the existing data away from dynamics that can be well-described by the mathematical model. Accordingly, the important information from Figure~\ref{Fig:PhenotypeExample} A) is the comparison 
\begin{align*}
G_{pheno}(\theta^*(\phi^i),\phi^i) \leq G_{pheno}(\Psi(\phi^i),\phi^i) < G_{pheno}(\theta^*(\phi^{i-1}),\phi^i),
\end{align*}
which demonstrates the accuracy of the continuation step~\eqref{Eq:ContinuationEqn} in driving a relative decrease in $G_{pheno}$.

Further, in Figure~\ref{Fig:PhenotypeExample} B), we show the cumulative number of objective function evaluations when calculating $\theta^*(\phi^j)$ for $j = 1,2,...,10$ when starting the optimization from $\theta^*(\phi^{j-1})$ and $\Psi(\phi^j)$. The total number of function evaluations used is lower when starting the optimization from the predicted MLE $\Psi(\phi^j)$ than when starting from $\theta^*(\phi^{j-1}$. More strikingly, the predicted MLE $G(\Psi(\phi^j),\phi^j)$ is comparable against $G(\theta^*(\phi^{j},\phi^j)$ in Figure~\ref{Fig:PhenotypeExample} A) and there is computational benefit to only calculating the predicted MLE $\Psi(\phi^j)$ rather than re-fitting the parameters. Taken together, the results shown in Figure~\ref{Fig:PhenotypeExample} demonstrate the accuracy and computation efficiency gained by calculating $\Psi(\phi^j)$.

We now demonstrate how to utilize the continuation frame work to identify additional time points to increase confidence in model parameters. We focus on the treated environment and consider additional time points $t_{s,i} = 3.1,3.2,3.3,3.4,3.5,5,7$ days with corresponding simulated measurements $\{ \phi_{i,s} \}_{i = 1}^7 = N(t_{s,i}).$ We perturb each of these simulated measurements by a fixed amount, $\Delta \phi = \pm 0.3 N(3.1) $, to give 14 additional, perturbed measurements. We appended each of these 14 measurements to the experimental data and predicted the MLE to these appended data sets. 

We calculated the relative change in the MLE for each model parameter and each of the 14 appended data sets. We note that each of the simulated data point occurs following the beginning of therapy. The immediate decrease observed in $N(t)$ following the beginning of treatment is due to the death of sensitive cells following treatment administration and controlled by the parameter $d_a^{max}$. From the biological interpretation of the parameters, we expect $d_a^{max}$ to be highly sensitive to perturbations in these data points. 

\begin{figure} [h!]
\includegraphics[scale=0.9, trim= 50 500 0 0,clip]{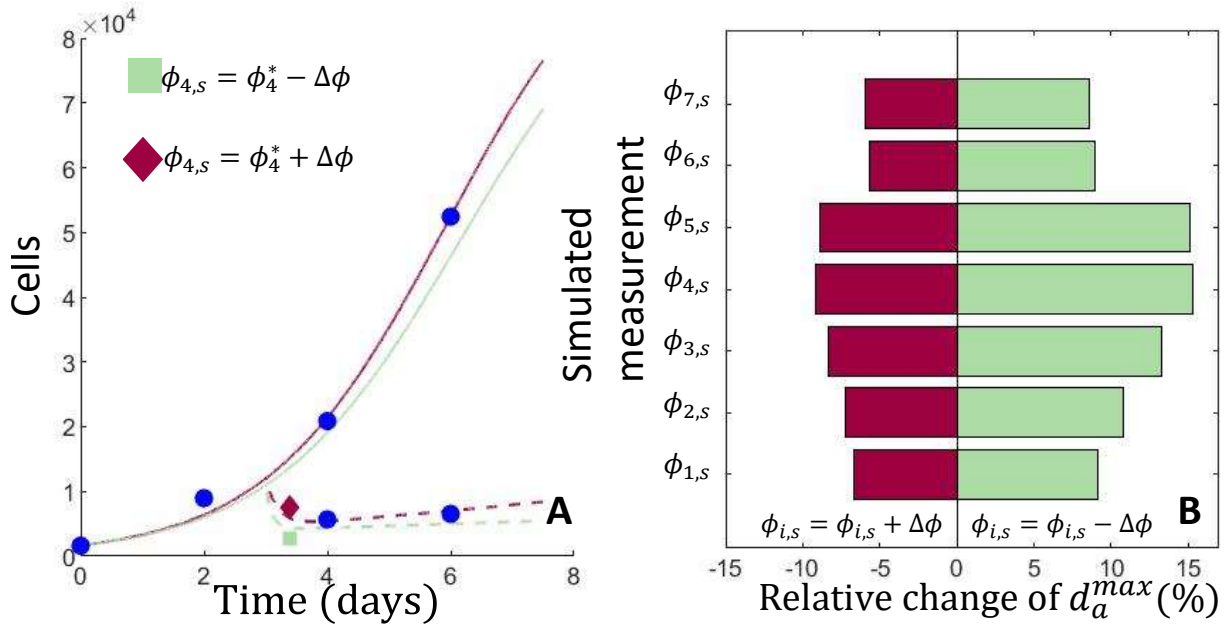} 
\caption{Evaluating additional time points to identify $d_a^{max}$ in an \textit{in vitro} model of NSCLC. Panel \textbf{A} shows the a selection of predicted model dynamics when fit to experimental data with a single additional time point $\phi_{i,s}^*$ that is perturbed by a $\Delta \phi$ from the true simulated value. For figure clarity, model trajectories corresponding to the perturbation of $\{ \phi_{4,s} \}$ is shown.  Panel \textbf{B} shows a tornado plot of the predicted relative change in the best-fit parameter $d_a^{max}$ for each additional simulated data point $\{ \phi_{i,s} \}_{i = 1}^7$. The left side of the tornado plot, in blue, shows the relative change when the perturbed value $\phi_{i,s} = \phi_{i,s}^* + \Delta \phi$ is larger than the simulated value $\phi_{i,s}^*$. The right-hand side, in orange, shows the relative change in $d_a^{max}$ when $\phi_{i,s} = \phi_{i,s}^* + \Delta \phi$ is smaller than the simulated value $\phi_{i,s}^*$.  }
\label{Fig:ExperimentDesignExample}
\end{figure}

As expected, $d_a^{max}$ was the most sensitive model parameter to perturbations of the simulated data and we show the percent relative change in $d_a^{max}$ from the unperturbed data in  Figure~\ref{Fig:ExperimentDesignExample} \textbf{B)}. As expected, the maximal death rate of sensitive cells increased when the simulated data point was decreased from the true value and decreased when the simulated data point was increased.  

The treatment sensitive population rapidly shrinks during therapy. The stabilization and rebound of the population during therapy is due to the expansion of the drug resistant population. This stabilization occurs once the drug sensitive population has been maximally suppressed which due to the drug effect. The most informative simulated data point, as measured by the magnitude of the relative change in the parameter $d_a^{max}$, was at time $t_{i,s} = 3.4$. At $t = 3.4$, drug sensitive cells are no longer dominant due to drug pressure. The depth of the population response to treatment, as measured by $N(3.4),$ is thus highly sensitive to death rate of these drug sensitive cells under treatment. In Figure~\ref{Fig:ExperimentDesignExample} \textbf{A)}, we show the simulated experimental measurements and predicted model dynamics for the most informative time point. The predicted model simulations capture the perturbed data point while retaining good fits to the true experimental data. 

\subsubsection*{Parameter continuation in a viral dynamics model}

The standard viral dynamics model has been extensively used to understand the dynamics of viral infection in HIV-1 \citep{Perelson2002}. The model tracks the concentration of uninfected target cells, $T(t)$, infected cells $I(t)$, and free infectious virus $V(t)$. Here, we follow \citet{Wu2008} and consider a model of HIV-1 dynamics where the target cells are CD$4^+$ T-cells. These cells are produced at a constant rate $\lambda$ and cleared linearly at rate $d$. Infection occurs at a rate $\beta$ following contact between a target cell and infectious viral particle and infected cells are cleared at rate $\delta$. Upon lysis, infected cells release $N$ viral particles into the circulation and free virus is cleared at a constant rate $c$. The viral dynamics model is given by
\begin{equation}
\left.
\begin{aligned}
\TimeDeriv T(t) & = \lambda -\beta T(t) V(t) -d T(t) \\
\TimeDeriv I(t) & = \beta T(t) V(t) -\delta I(t) \\
\TimeDeriv V(t) & = \delta N I(t) - c V(t).
\end{aligned}
\label{Eq:StandardViralDynamics}
\right \}
\end{equation} 
 It is common to set $p = \delta N$ so the final equation for $V(t)$ becomes 
 \begin{align*}
 \TimeDeriv V(t) & = p I(t) - c V(t),
 \end{align*}
and the system~\eqref{Eq:StandardViralDynamics} is equipped with initial conditions $T(0) = T_0, I(0) =I_0, $ and $V(0) = V_0$. In typical clinical studies, temporal data is only collected for circulating free virus so the model output corresponding to the calibration measurements is 
\begin{align*}
y_i (\theta) = \log_{10} (V(t_i,\theta)), 
\end{align*}
where using $log_{10}$ measurements of viral load is standard in HIV studies. 

During antiretroviral therapy (ART), the viral load may fall below the limit of detection of standard assays. While there are a number of techniques to account for this censored data, we do not consider data collected during ART, so the objective function is given by the sum of squares error
\begin{align}\label{Eq:ViralDynamicsObjective}
G_{HIV}(\theta,\phi) = \sqrt{ \displaystyle \sum_{i=1}^n \left( \log_{10}(V(t_i,\theta) - \log_{10}(\phi_i) \right)^2 }. 
\end{align}
\citet{Wu2008} characterized the identifiability of this model using a higher order derivative method. They found that, if the initial conditions of the model $T_0,I_0,$ and $V_0$ are known, then all six model parameters $\theta = \{ \beta, d, \delta,c, N, \lambda,\}$ are identifiable. To illustrate their results, they fixed $\theta = \{  (2 \times 10^{-5}, 0.15, 0.55, 5.5, 900, 80\}$ and simulated the ODE model \eqref{Eq:StandardViralDynamics}. They sampled the simulated viral load at $37$ distinct time points and added noise $\epsilon_i$ sampled from a Gaussian distribution with  $\mu = 0$ and $\sigma^2=1$ \citep{Wu2008}.

 In Section~\ref{Sec:PhenotypeExample}, we demonstrated the effectiveness of our continuation technique by focusing on objective value function and computational efficiency in calculating the MLE. Here, we illustrate how model dynamics evolve during MLE continuation.  We follow \citet{Wu2008} but consider a smaller subset of calibration data collected at time $t_i =  \{ 0.4, 1, 8, 14, 20, 36, 46, 58\}$. We add noise $\epsilon_i^0$ sampled from a Gaussian distribution with $\mu = 0$ and $\sigma^2 = 0.15$ so the initial calibration data is 
\begin{align*}
\phi_i^0 = \log_{10} (V(t_i,\theta)) + \epsilon_i^0. 
\end{align*}

We first fit the model to the simulated data $\phi_i^0$ to obtain an initial MLE. We then generate 4 additional viral load time courses $\{ \phi_i^j\}_{j=1}^{10}$ by 
\begin{align*}
\phi_i^j =  \phi_i^{0} + h_{step} |\epsilon_i^{j}| 
\end{align*}
for $\epsilon_i^j $ sampled from a Gaussian distribution with $\mu = 0$ and $\sigma^2 = 1$ and $h_{step} = \pm 0.1, \pm 0.2$. This collection of $4$ data sets could feasibly represent experimental data measured from an increasingly large sample drawn from a population of HIV-1 positive individuals with population viral dynamic parameters given by $\theta = \{  (2 \times 10^{-5}, 0.15, 0.55, 5.5, 900, 80\}$. Here, we test the ability of our continuation technique to predict reasonable viral dynamic curves \textit{without} refitting the data. 

\begin{figure} [h!]
\includegraphics[scale=0.9, trim= 50 590 0 0,clip]{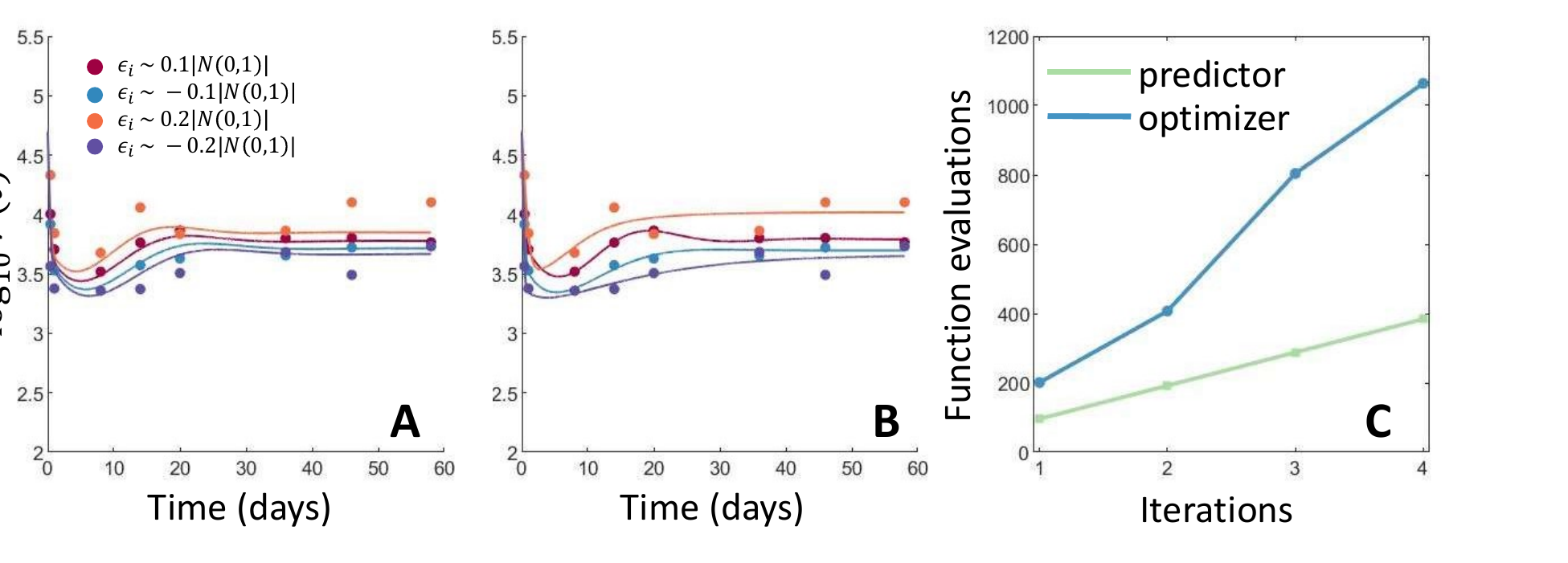} 
\caption{  Comparison of predicted model fits to randomly perturbed data. Panels \textbf{A} and \textbf{B} show model trajectories obtained using predicted model parameters to the simulated experimental data perturbed by $\phi_i^{j} = \phi_i^{0} + h_{step} |\epsilon_i^{j}| $. Panel \textbf{A} shows the predicted model fits to the experimental data  while \textbf{B} shows the model fits to data resulting from the true MLE. Panel \textbf{C} shows the number of objective value evaluations required to predict the MLE using this continuation technique or fit the model parameters to the perturbed data using the known parameters as a starting guess. }
\label{Fig:ViralDynamicsExample}
\end{figure}

In Figure~\ref{Fig:ViralDynamicsExample} A), we compute the predicted $\Psi(\phi^j)$ and plot the predicted model dynamics obtained from $\Psi(\phi^j)$ against the perturbed data $\phi^j$. In Figure~\ref{Fig:ViralDynamicsExample} B), we show the fit model predictions to the perturbed data. In each case, the viral dynamics show comparable model predictions for the fit and predicted model parameters demonstrating that our continuation method can successfully predict reasonable model simulations. In fact, the Bayesian Information Criteria \citep{Kass1995} indicates no significant differences between the predicted and true MLE for all 4 data sets. However, Figure~\ref{Fig:ViralDynamicsExample} C) shows the significant computational improvement obtained by only calculating the continuation step rather than fitting all model parameters at each step. The predicted model dynamics track the true viral load trajectory. 

\begin{figure} [h!]
\includegraphics[scale=0.9, trim= 50 480 0 0,clip]{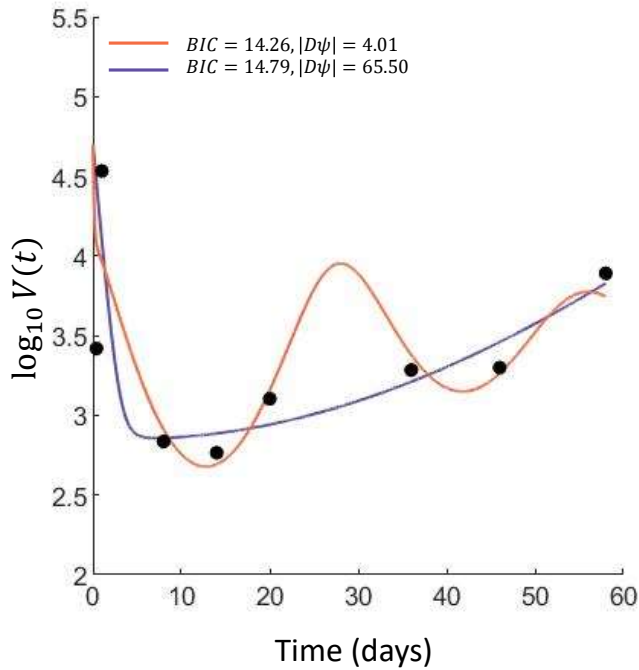} 
\caption{ Comparison of two potential fits to randomly perturbed viral dynamics models. Model trajectories obtained from two local minima from fitting 10 initial guesses to viral load data shown in black. Both trajectories accurately describe the viral load dynamics as evidenced by a small difference in BIC. However, the parameter estimate corresponding to the oscillatory trajectory is much more robust as measured by $\|\Diff \Psi( \phi^1)\|$.  }
\label{Fig:MLERobustnessFigure}
\end{figure}

It is common to find numerous local minima of \eqref{Eq:ViralDynamicsObjective} when fitting \eqref{Eq:StandardViralDynamics} to simulated data. As measured by the value of the log-likelihood function or information criteria, these local minima can produce comparable fits to a given data set despite different dynamics. We perturbed the initial data set $\phi_0$ by \begin{align*}
\log( \phi_i^1) =  \log(\phi_i^{0}) +   0.8 \epsilon_i 
\end{align*}
for $\epsilon_i$ sampled from a Gaussian distribution with $\mu = 0$ and $\sigma^2 = 1$. We fit this perturbed data from 10 distinct initial guesses using \texttt{fmincon} \citep{Matlab2017}. These 10 starting initial guesses converged to two local minima. We denote the corresponding parameter estimates by $\hat{\theta}_1$ and $\hat{\theta}_2$ and plot the resulting model trajectories in Fig~\ref{Fig:MLERobustnessFigure}. These fits are indistinguishable by BIC and both appear to accurately describe the viral load data. Consequently, it is not obvious which of $\hat{\theta}_1$ and $\hat{\theta}_2$ best describe the data.  

However, it is reasonable to expect that the MLE should be robust to small perturbations of the calibration data. We measure the robustness of each of these minima by calculating $\|\Diff \Psi( \phi^1)\|$ at $\hat{\theta}_1$ and $\hat{\theta}_2$. A smaller norm $\|\Diff \Psi( \phi^1)\|$ implies less sensitivity of the MLE to perturbations of the calibration data. For the example shown in Fig~\ref{Fig:MLERobustnessFigure}, there is a 16 fold difference in sensitivity to calibration data. In this way, $\Diff \Psi$ can be used to distinguish between otherwise similar fits.  We suggest that, when choosing between multiple fits with similar BIC values, the parameter estimate with the smaller sensitivity to the data is a more robust, and thus preferential, fit.  

\section{Discussion}

Parameter fitting is crucial step when using mathematical models to predict novel treatment strategies, extrapolate from clinical trials, identify new drug targets or schedules, or propose non-pharmaceutical interventions \citep{Brady2019,Cassidy2020b,Cassidy2019}. However, parameter fitting can be difficult and computationally expensive. A large variety of fitting techniques have therefore been developed to calibrate model predictions against data \citep{Toni2009,Horbelt2002,Kreutz2013,Lauss2018}. Moreover, mathematical modeling is increasingly applied to understand emerging data and make real-time predictions. In this case, as new data emerges, the model parameters must be refit with potential computational cost. Here, we developed a continuation type technique to quantify how updates to experimental data will impact the MLE and predict the evolution of the MLE as a function of the experimental data used to calibrate the model. 

We used the implicit function theorem to calculate the trajectory of the MLE through parameter space. As the implicit function theorem only guarantees the existence of a differentiable trajectory $\Psi$ through calibration data--parameter space, we utilized the first order Taylor expansion $\Psi$ to extrapolate the evolution of the MLE due to changes in experimental data. We showed how this calculation is intrinsically linked to local sensitivity analysis and the curvature of the objective function. In two examples drawn from mathematical biology, we showed how this continuation technique can predict acceptable model fits to experimental data while significantly reducing computational overhead. In fact, in most applications, our continuation technique requires no dedicated computational overhead as the Hessian of the objective function is calculated at each step when using common optimization algorithms, such as \texttt{fmincon} \citep{Matlab2017}, and local sensitivity analysis is a standard step in model fitting. 

Perhaps more importantly that gains in computational efficiency, our approach explicitly identifies relationships between individual experimental measurements and parameter estimates. Our approach addresses similar questions to local sensitivity analysis from a distinct perspective. Rather than using simulations to understand how small perturbations in model parameters from the best-fit parameters change model outputs as in standard sensitivity analysis, we quantify how changes in the training data impact the best-fit parameters and measure the sensitivity of the best-fit parameters to variations in this calibration data. As we showed in Section~\ref{Sec:PhenotypeExample}, this perspective can be used to suggest additional experimental measurements to increase confidence in model parameterization. Further, we showed how to use $D\Psi$ to understand which experimental measurements are most informative for model parameterizations and identify redundant measurements that do not provide additional information for parameter estimation.  

Our technique is a type of local analysis that explores the functional dependence of the MLE on experimental data starting from a pre-identified MLE. Specifically, we assume that the Hessian of the objective function is invertible at the MLE and our results are necessarily local in parameter space as we are extrapolating from a pre-identified MLE. Nevertheless, our examples show the utility of our continuation approach for even large perturbations of the experimental data. 

Despite these limitations, we developed a continuation-type technique to predict the functional dependence of a MLE on the experimental data used to train a mathematical model. While we have focused on applications in mathematical biology, our approach is immediately portable to other domains. As our method is independent of the number of data points, our approach could be particularly useful in big-data applications. Ultimately, our results offer a unified approach to quantify the relationship between training data and best-fit model parameters and to leverage this understanding to suggest additional experiments to increase confidence in model parameterization.

\section*{Data access statement}
The code and data underlying the results in this manuscript is available at \texttt{https://github.com/\allowbreak ttcassid/MLE$\_$Continuation}.


\end{document}